# CRYPTOCURRENCY, SANCTIONS AND AGRICULTURAL PRICES

An empirical study on the negative implications of sanctions and how decentralized technologies affect the agriculture futures market in developing countries.

Agni Rajinikanth

International Economics

**Abstract**

This research paper focuses on the implications of cryptocurrency to fight and evade sanctions. The 2022 Russia Ukraine War has led to many sanctions being placed on Russia and Ukraine as well as the nearby locale. The paper will discuss the impact the 2022 Russian Sanctions have on agricultural food prices and hunger. The paper also uses Instrumental Variable Analysis to find how Cryptocurrency and Bitcoin can be used to hedge against the impact of sanctions. The 6 different countries analyzed in this study include: Bangladesh, El Salvador, Iran, Nigeria, Philippines, and South Africa, all of which are heavy importers of wheat and corn. The paper shows that although Bitcoin may be volatile compared to other local currencies, it might be a good investment to safeguard assets since it is not correlated with commodity prices. Furthermore, the study demonstrates that although transaction counts per day don't have a strong relationship with prices, transaction volume does have a strong relationship.

**Introduction**

The 2022 Russia Ukraine conflict has garnered widespread international attention, with many international players taking on strong stances in the war. Bodies such as the United States, UK, and the EU have all condemned Russia by laying economic sanctions upon the nation's imports and exports. Russia and Ukraine are major exporters of the global food and oil supply. These embargos have stifled international trade causing commodity prices to increase dramatically and creating a humanitarian crisis. Wheat and Corn are both staple food sources grown in these regions together amounting to around 66% of the global population's diet. These crops are stable goods necessary for survival. Around 30% of the global production of wheat and 15% of the production of corn comes from both Russia and Ukraine, which affects these prices dramatically[1]. Production is also affected by the dramatic shocks in oil prices, which increases the cost of goods transportation and causes the supply chain issues. Fertilizer, which is the backbone of the food supply, is also a major exported product from Russia increasing food prices even further. When these agricultural commodities become overpriced, developing countries and low-income households are impacted the greatest, leading to hunger and welfare decline.

Although these sanctions are meant to be targeted towards the country's governments and elites, innocent citizenry are affected the most as collateral damage from this economic warfare. These governments and elites have capitalized on cryptocurrencies, a developing technology, which allows parties to anonymously transfer money without any intermediary helping them evade much of the impacts of sanctions[2]. This fosters illegal activity such as sanction evasion through a series of protocols. With the increasing use of cryptocurrency in the world, many rich elites are escaping the impact of these embargoes making the common folk of society carry the bulk of the burden.

---

[1] https://oec.world/en/profile/hs
[2] https://jbonneau.com/doc/BMCNKF15-IEEESP-bitcoin.pdf

      The UN's World Food Programme has identified a dramatic rise in global hunger malnourishment and poverty because of the rising food prices. The United Nations blames the Russia Ukrainian war conflict for this global problem causing heightened prices as a result of the Covid-19 Pandemic to increase much further. The number of malnourished children has increased drastically by 13 million from 2021 to 2022. The barriers placed on trade because of these embargoes has threatened the global food supply. The problem has caused other agricultural commodity exporters such as India and Argentina to withhold their exports to satisfy demand for their own populations affecting global supply further. While this problem may only pose a minor inconvenience for the rich class of society who can easily convert assets and mitigate the problem, Low-income people as well as those in developing countries struggle the most during these times. When many people are unable to have proper food and nutritional meals, worldwide goals and progress are set back, which leads to major humanitarian crisis. In some countries these rising prices lead to currency inflation because of government assistance towards those affected which devalues the national currency and exacerbates hunger. If the rich can exploit cryptocurrency technologies to evade sanction measures, common citizens should also be able to hedge against the impacts of the rising agricultural prices. As cryptocurrencies have become more and more prevalent in society, it is important to explore if adopting cryptocurrency can help civilians prepare for the rising agricultural prices and guard against the negative implications of sanctions.

**Background**

**Sanctions on Russia**

Many nations have been employing sanctions as tools to punish opponents and enemies without the direct escalation of war making sanctions an important role in international relations especially in peacekeeping. But before the 1980's, sanctions were not a popular tool for international diplomacy and were not widely used in cross country relations. But with the rise of the United Nations and other regional organizations such as the European Union, sanctions have become more popular. As these international bodies grow in membership, political influence, and power the use of sanctions is often their go-to strategy for punishing entities which threaten international peace and security.

One of the earliest known examples of sanctions occurred in 432 B.C.E when Athens prohibited importing and selling of products from Megara in retaliation for a kidnapping. During the American Revolution, American patriots adopted mass boycotts to pressure the British Empire to repeal its taxation policies. But this type of sanctions dramatically changed during the world wars when states prohibited trade, canceled contracts, and confiscated goods of all enemy countries. After World War II, the United Nations adopted new measures that allowed the security council the ability to enact economic and diplomatic sanctions, which created a spree of embargos in the 1990s. This period led to a shift in sanction policy after the 1990 Iraqi sanctions led to a humanitarian crisis. The negative effect led to reform and brought up the discussion of the ethicality of sanctions as counteractive measures. As witnessed by the sanctions on Iraq, sanctions not only hurt the political elites which the sanctions are meant to target but also the common people of the land. Although slight reforms have been made the problem still exists and is a major issue to this date.

Russia has been one such country that has been targeted by sanctions from countries around the globe repeatedly throughout recent years. These sanctions on Russia are largely autonomous, implemented by individual countries usually without the collaboration within a central body such as the United Nations. In 2008, 2014 and 2022 Russia has dealt with sanctions due to its invasions, annexations, and occupation of nearby territory. These sanctions have been placed in addition to asset freezing demonstrating economic warfare. The primary drivers of these sanctions have been the United States in cooperation with the European Union, who have levied these sanctions in response to the aggression and invasion Russia has shown towards Ukraine, Crimea, Georgia, and the nearby territory[3]. In 2022, Russia attempted to extend its control over Ukraine by sending in troops. Russia claims that NATO and its plans to potentially include Ukraine within their organization is a threat to its national security. Russia also has completely recognized two regions in Donbas region as part of Russia because of its overwhelming pro-Russia support. On the other hand, Ukraine and its people have come to the streets to protect their homeland and fight the Russian Invasion. Although a unified force of sanctions was not able to take place during the 2014 annexation of Crimea the 2022 invasion of Ukraine has brought forth a more unified front among these different bodies. In retaliation to the sanctions imposed, Russia had enacted their own boycotts of products attempting to stifle and deter the pursuits of their antagonists. Russia and Ukraine are heavy exporters of commodities such as oil, wheat, corn, and fertilizer. Coupled with disruptions in payment systems, shipping, and trade, commodity prices have soared. These sanctions and embargoes lead to a rise in food prices, which has affected low-income communities and developing countries and removing their access to these vital agricultural resources.

---

[3] https://crsreports.congress.gov/product/pdf/R/R45415

**Future Markets and Agricultural Prices**

The futures market is built upon the premise of trading goods at a certain price in the future involving two parties both a producer and consumer. The dynamics of this market allows those who don't want to buy the product but rather wish to only make financial profit to take part in the trading. Most exchanges allow traders to offset their positions and opt for a cash settlement instead of a physical good. Therefore, futures markets are often traded in two different methods physically with goods or just with money instead. In addition, the futures price is determined based on supply and demand principles and considers the theory of price storage to help determine the most accurate prices[4]. Furthermore, the futures market correlates to the spot price of the commodity which can be calculated using the spot price parity theorem. Overtime, the futures price converges around the spot price therefore making the futures market price a good indicator of the commodity's worth[5]. The futures market is essential because it allows parties to reduce risk in their investments which helps benefit both the produce and consumer of the product. Players use the futures market to hedge against any losses in the future such as increased prices, and natural disasters, among other occurrences.

The creation of the World Trade Organization has spurred growth in commodity markets, making it what we see today. The organization has brought much more attention towards the system to encourage trade and prosperity. But there are some occurrences in which these markets can be plagued with high volatility with some drawbacks. Speculators can dramatically alter the market by pumping in money into the exchange and trading with only financial incentives in mind. These speculators pump tons of money into the market leading to an uptick in liquidity. This sometimes leads to increased volatility in the market making prices unpredictable which can be detrimental especially for farmers. The 2007-2008 housing market crash is one similar example in which speculators played a major role in bringing the market crash. Instances such as these cause markets to crash and mostly affects the have-nots of society.

---

[4] https://www.jstor.org/stable/1816601?seq=1
[5] https://www.sciencedirect.com/science/article/pii/0927539895000143

International demand for food has risen as the global population has increased dramatically over the past few years. The global population is currently approximately 8 billion people with experts predicting it will reach around 10 billion by 2050. With the rise in population also comes the need for increased food production supply to meet the growing demand. International trade has also introduced new products to the world which homogenized the world food diet. This has put a strain on certain staple crops such as wheat and maize[6]. The Food and Agricultural Organization estimates that around 2/3 of the world's energy intake comes from the staple crops wheat and maize. Furthermore, the same study shows that 80% of the imports of these commodities go to developing countries. To cope with the rising populations, fertilizer has also been a tool used to increase crop production to meet this increasingly global demand. But as Russia and Ukraine are heavy exporters of these commodities, prices as well as futures markets have been heavily affected.

Most of the wheat and corn from this region goes to developing countries who have been heavily impacted by the rising prices. oil and fertilizer trade have also been impacted causing prices to soar as well. Much of the shipment and trade of these commodities occurs through the black sea which has faced halts and blockades due to the ongoing war. With prices already soaring due to inflation and Covid-19, the problem has worsened drastically. All these commodities play a major role in the futures market; their price impacting the world hunger heavily.

---

[6] https://pubmed.ncbi.nlm.nih.gov/17307959/

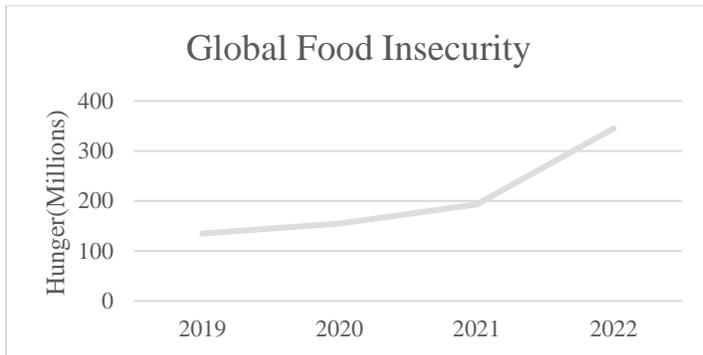

*Fig. 1. Global Food Insecurity*

**Cryptocurrency**

Recently, there has been great hype and interest surrounding the new virtual decentralized technology, cryptocurrency. These new currencies are independent of many national currencies such as the dollar and operate without any central authority or figure. This feature makes the monetary transfer decentralized which allows the transfer to be quick and easy. This global currency has recently been increasingly adopted all over the globe, with countries such as El Salvador adopting Bitcoin as legal tender. Other countries have used cryptocurrency to establish and promote their own local virtual coins. Globally, user adoption has soared rising over 880% within the past year, increasing usage especially in new emerging markets and developing countries.

The widespread adoption of cryptocurrencies allows it to become a truly global international monetary system which directly competes with national currencies[7]. Because cryptocurrencies don't have a central authority, they are less impacted by inflation and deflation making, them a highly anticipated investment. Furthermore, the ease in transactions which cryptocurrencies provide allows them to be a widely adopted method for sending money between countries. Research from Binance shows that the most popular cryptocurrency is Bitcoin, which around 65% of crypto users own. The same research also shows that 63% of users use their disposable funds when investing in cryptocurrency. Most crypto users are from the upper and higher educated class of society, showing the inequality in accesses to this technology. The technology has allowed 52% of users to generate stable revenue from the currency proving its efficiency and concept model. This growing interest has caused the price of Bitcoin to increase more than 540,000% from 2012 to 2020 and it is projected to grow by 56.4% annually in the coming years[8]. The primary reasons for purchasing crypto vary and range from mistrust in the government to long term investment opportunities. These changes have encouraged businesses to accept crypto as payment as well because of its benefits in sales and attracting customers.

Because of the anonymity cryptocurrencies provide, they have also become a haven for illegal activities and fraud. Although illicit activity accounts for only 2.1% of all cryptocurrency transactions, some have criticized it for fostering illegal trade, money laundering and frauds. Over $14 billion worth of cryptocurrency has been traded with illicit addresses which is up from 7.8

---

[7] https://go.chainalysis.com/rs/503-FAP-074/images/Geography-of-Cryptocurrency-2021.pdf
[8] https://research.binance.com/static/pdf/Global_Crypto_Index_2021.pdf

billion in 2020. These concerns have also split into the topic of sanctions with concerns on how sanctioned countries can escape their punishments through cryptocurrency.

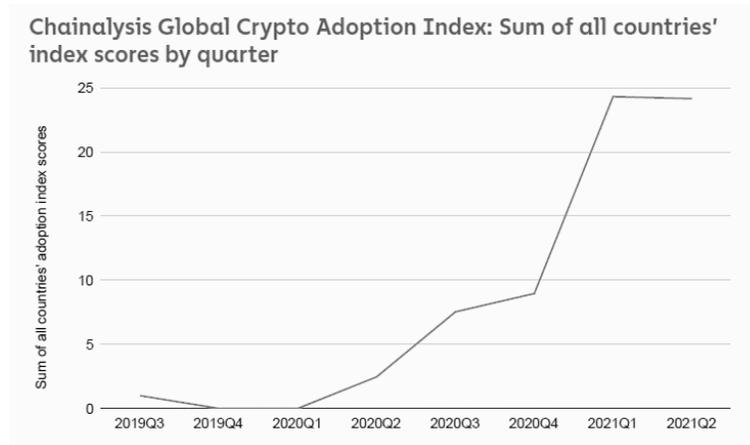

*Fig. 2. Chainalysis Global Crypto Adoption Index*

**Literature Review**

Studies have been conducted evaluating the efficacy of sanctions and the impact they have on affected countries. Drezner(2010) has showed the ineffectiveness of sanctions in their ability to deter opposing nations. When a combined force of UN sanctions was placed in Iraq in the 1990s, despite the multiple hardships Iraq faced they did not waver. Even though sanctions were placed on for an incredible 8 years, Iraq did not budge while inflation and mortality levels rose. Drezner points out that even though sanctions are not effective and have backlashes, they continue to be important tools of foreign policy[9].

Hinz and Monastyrenko (2022) showed the effect the 2014 sanctions on Russia had on the regional economy. This paper focuses on the impact of a self-imposed embargo that Russia placed during the 2014 sanction and annexation of Crimea. The Russian embargo increased the prices of the commodities which were embargoed. This price hike would remain stagnant for 2 years and cause welfare declines for Russian citizens. The study shows that a self-imposed embargo has the capability to cause self-harm to the nation[10]. Crozet and Hinz (2020) discuss the cost Russian Sanctions paved for the imposing countries. The study found that the sanctioning countries lost $43 billion regarding exports because of sanctions. This therefore shows that there are also negative repercussions of sanctioning[11]. Boulanger et al. (2016) shows the impact the 2014 Russia Sanctions had on food prices including imports and exports. In this episode, Russia imposed bans on agricultural produce from external countries including the EU which caused heavy losses for them. The study is also notable at showcasing that the EU had minimal losses compared to the Russians because they were able to find alternate sources to purchase their produce.

---

[9] https://go.chainalysis.com/rs/503-FAP-074/images/Crypto-Crime-Report-2022.pdf

Kuzminoz et al. (2018) describes in their research the growing market share Russia continues to gain in the agricultural industry. Although the industry lacks proper agricultural technology and suffers from productivity, because of high self sufficiency rates, Russian and regional economies to increase have increased their produce and expanded into new and developing markets[12]. Dreze and Gazdar (1992) analyze the impact of the economic sanction on Iraq have found that the economic sanctions caused high levels of mortality, poverty, and malnutrition. Due to rise in prices, local welfare declined resulting in widespread famine and poverty[13]. Klomp (2020) shows that the news regarding the Russia Sanctions triggered a reduction in the returns in the commodity futures price effecting on the futures market. These research studies allow us to analyze the negative repercussions of sanctions on food prices and its contribution towards hunger.

---

[10] https://www.cambridge.org/core/books/the-sanctions-paradox/4542E89CDBABCBE49039C580F9A7F5F3
[11] https://www.sciencedirect.com/science/article/pii/S0022199622000137#!
[12] https://onlinelibrary.wiley.com/doi/full/10.1111/1477-9552.12156
[13] https://onlinelibrary.wiley.com/doi/full/10.1111/1746-692X.12184

Busch and Tierno(2022) analyze the effects cryptocurrency have on fraud and illicit activities. They discuss the concerns that Russians may user cryptocurrencies to evade sanctions through its anonymity. Potentially they could use chain hopping techniques, un-hosted wallets, and peer to peer exchanges, all of which make traceability difficult. The report concludes that evasion is possible but not on large scale endeavors due to liquidity issues[14]. Vendier (2020) also shows that many countries are attempting to explore cryptocurrency to counter the impact of sanctions. Countries such as North Korea, Iran and Russia have explored these measures as the technology is growing in capabilities. These countries devalue the sanctioning country's power and currency by performing this evasive measure.

Dudley et al. show that cryptocurrencies can be leveraged to circumvent the financial blocking sanctions are meant to impose. This leads to obstacles in national security and lack of adequate regulation and allows for the malign activity to continue[15]. Graur et al. (2022) Global Crime Report shows that sanctioned individuals in Russia account for most of the cryptocurrency activity. The research also indicates that other countries like Iran, who are also sanctioned, employ their resources in Bitcoin Mining to evade the sanctions because their energy prices are low[16].

Most of the research done in this field surrounds how governments on a large scale can evade the impact of sanctions. With the research available showing the impact sanctions have on the food prices and later hunger, it is important to discuss the role of sanctions. Sanctions affect common people as collateral damage when they target the elites and governments. These elites and governments can escape the bulk of the punishments through cryptocurrency leaving the common people to face a lot of burdens. Therefore, this study explores how common people, usually in developing countries, can use cryptocurrencies to help fight the rise in agricultural prices during periods of sanctions.

---

[14] https://www.sciencedirect.com/science/article/abs/pii/0305750X9290121B
[15] https://crsreports.congress.gov/product/pdf/IN/IN11920
[16] https://ndupress.ndu.edu/Portals/68/Documents/jfq/jfq-92/jfq-92_58-64_Dudley-et-al.pdf

## Data

**Countries**

To determine whether citizens of developing countries are affected by sanctions and can use cryptocurrency to hedge against the rising food prices. The 6 countries chosen are Bangladesh, El Salvador, Iran, Nigeria, Philippines, and South Africa. The key parameters which influenced choosing these countries included demographics, agricultural production, cryptocurrency, and international trade. The International Monetary Fund classifies all 6 of these countries as developing nations. This category is important because any fluctuation in prices affects these countries more than others making them more susceptible to dramatic fluctuations. Being heavy importers of crops such as wheat and maize, these countries are affected heavily when prices increase because they are staple foods which many depend on for their livelihoods. These countries are also growing in cryptocurrency ownership and are relatively high in adoption compared to other similar countries. Finally, all these 6 countries have futures market exchanges in their locale where they trade and barter. Although the usage of these markets differs, the necessary infrastructure is in place for the markets to grow in the future. By fitting these constraints, these countries make excellent samples to study for this research paper. The graphs below are food prices obtained from the Food Agricultural Organization and Interest Rates obtained from each respective country's national bank report.

### Bangladesh

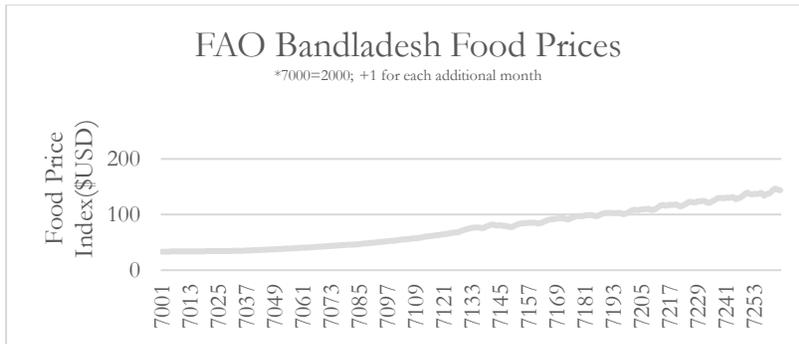

*Fig. 3. FAO Bangladesh Food Prices*

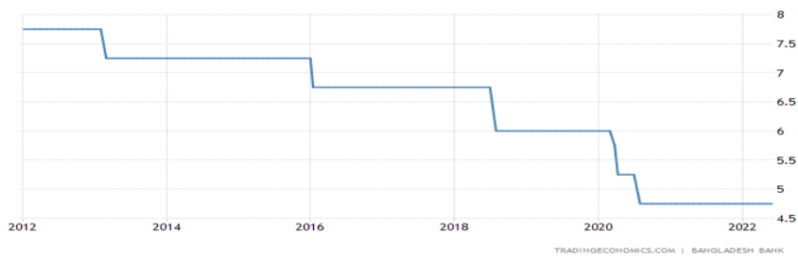

*Fig. 4. Bangladesh Interest Rates*

### El Salvador

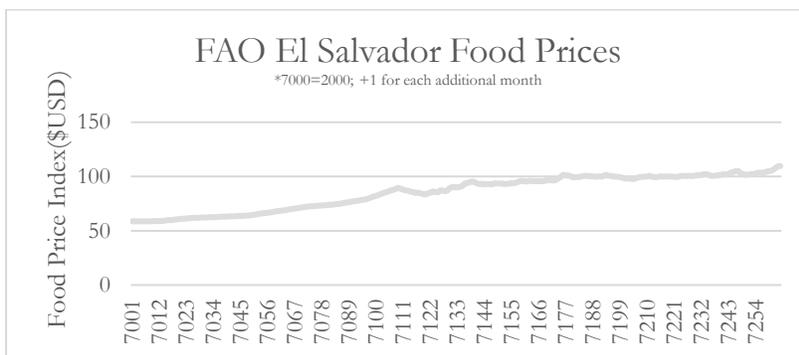

*Fig. 5. FAO El Salvador Food Prices*

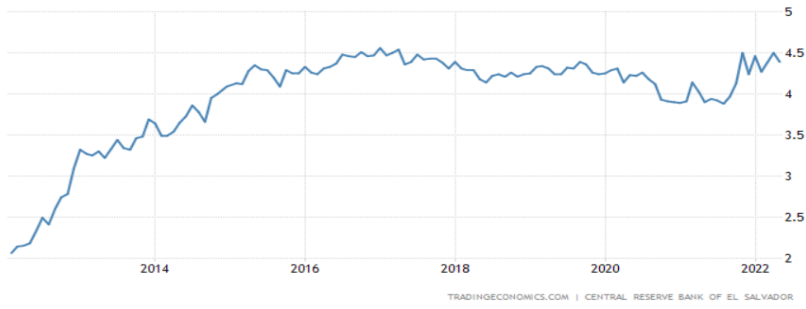

*Fig. 6. El Salvador Interest Rates*

**Nigeria**

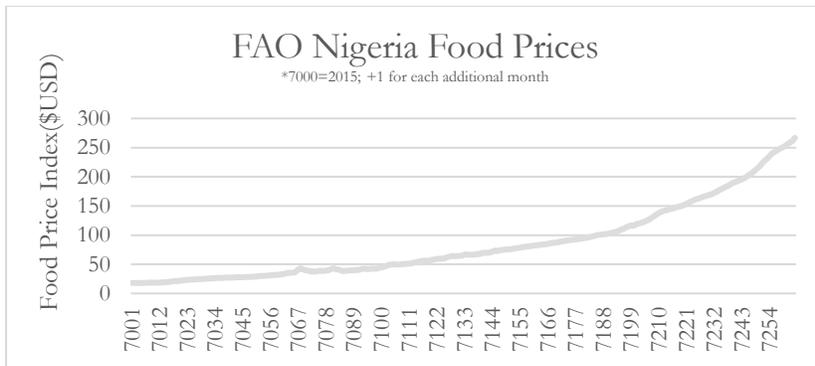

*Fig. 7. FAO Nigeria Food Prices*

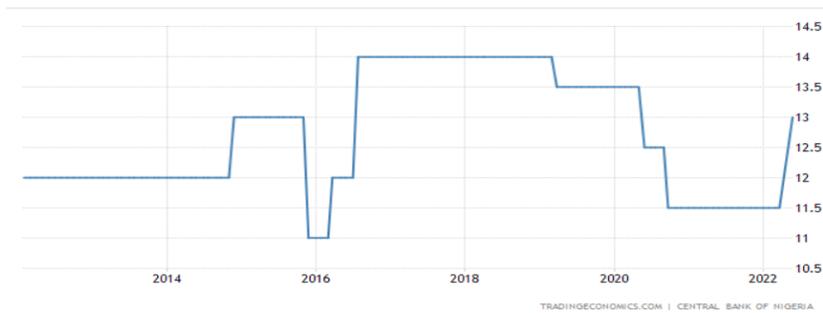

*Fig. 8. Nigeria Interest Rates*

**Iran**

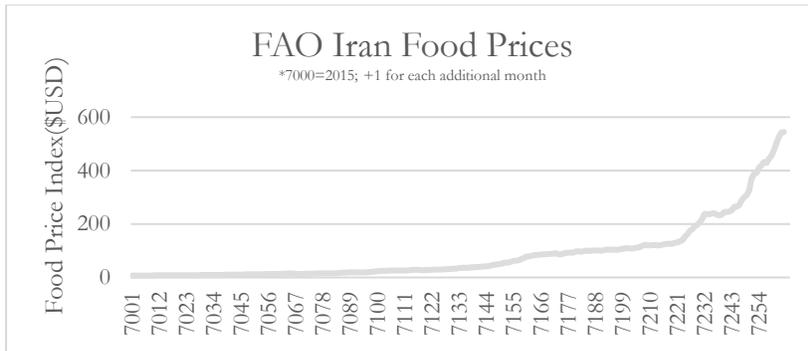

*Fig. 9. FAO Iran Food Prices*

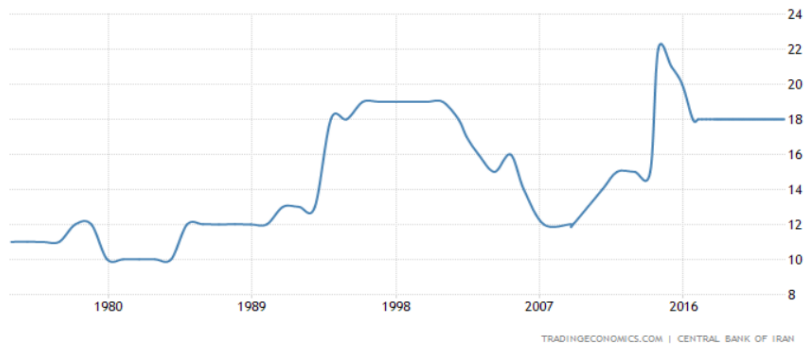

*Fig. 10. Iran Interest Rates*

**Philippines**

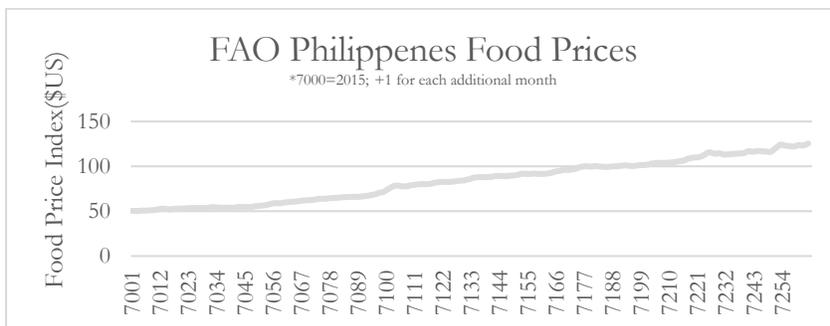

*Fig. 11. FAO Philippines Food Prices*

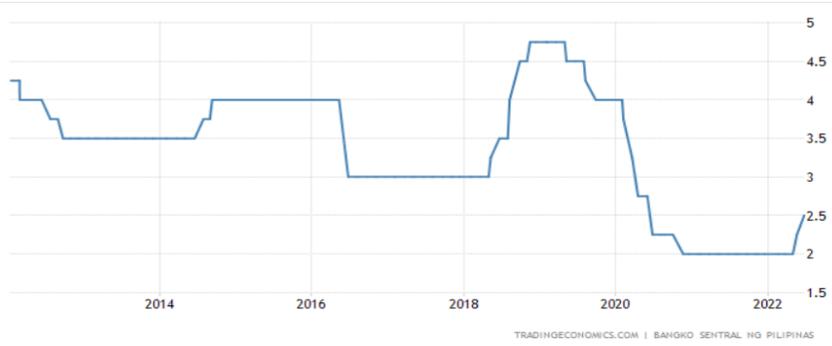

*Fig. 12. FAO Philippines Interest Rates*

**South Africa**

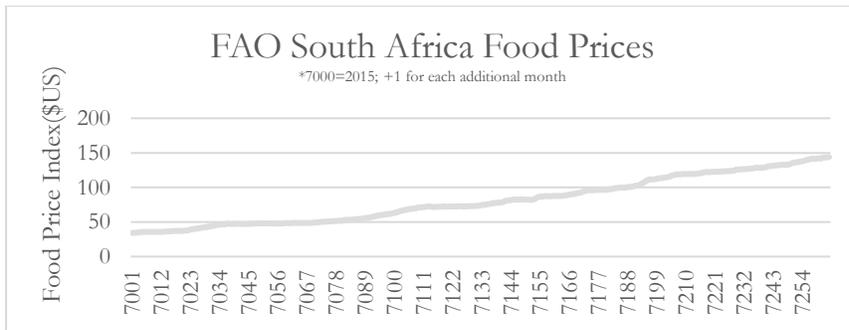

*Fig. 13. FAO South Africa Food Prices*

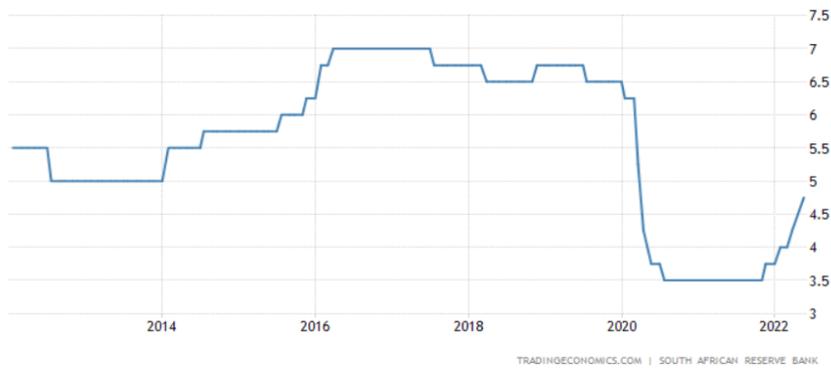

*Fig. 14. FAO South Africa Interest Rates*

**Analysis**

As we can see from the charts above, the FAO Food prices Indices show that general food prices have all increased because of the sanctions from Russia. Food prices have all increased in the 6 countries according to statistics from the Food Agricultural Organization. Furthermore, the second graphs under each country shows the interests rates of each country which directly correlates to inflation in the country according to the Quantity Theory of Money. During times of dramatic price hikes and humanitarian crises such as Hunger, governments pump and print more money into the economy. Interest rates and Inflation are inversely proportional to each other. In Nigeria, Philippines and South Africa interest rates declined then increased in 2022. Bangladesh and El Salvador's Interest rates varied opposite to each other; Bangladesh decreased while El Salvador increased. Not enough data was provided to make proper conclusions for Iran. With this knowledge, we can see the different approaches governments take to solve the crises and how food prices correlate with interest rates as well.

**Commodity Crop Prices**

The following graphs show the changes in prices of the two main agricultural commodities Wheat and Corn during 3 different episodes of sanctions throughout Russia's History. The data was obtained from the World Bank Commodity Dataset and includes worldwide price points per month. During all three episodes, sanctions were placed by external countries on Russia although the magnitude and impact of the sanctions wavered during each episode.

**2008 Invasion of Georgia**

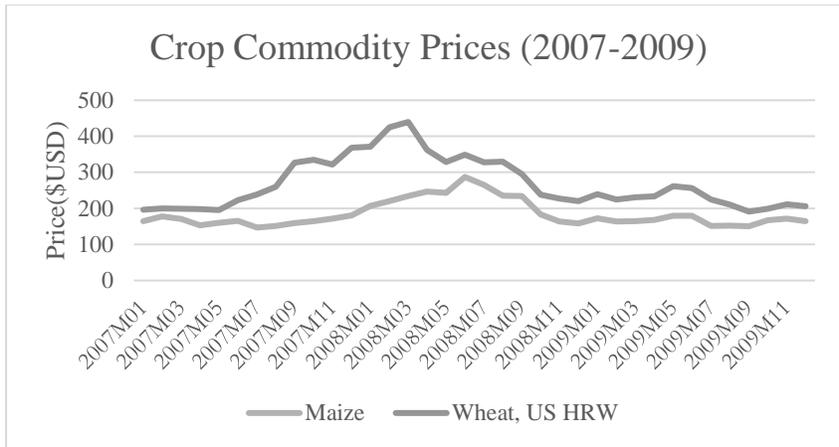

*Fig. 15. Crop Commodity Prices (2007-2009)*

**2014 Annexation of Crimea**

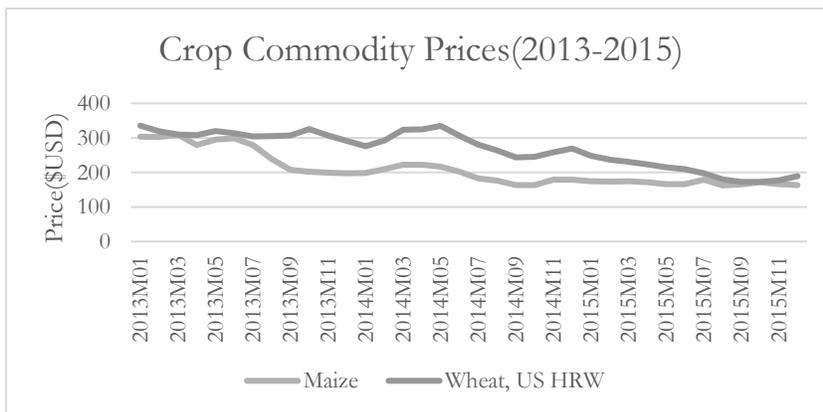

*Fig. 16. Crop Commodity Prices (2013-2015))*

**2022 Invasion of Ukraine**

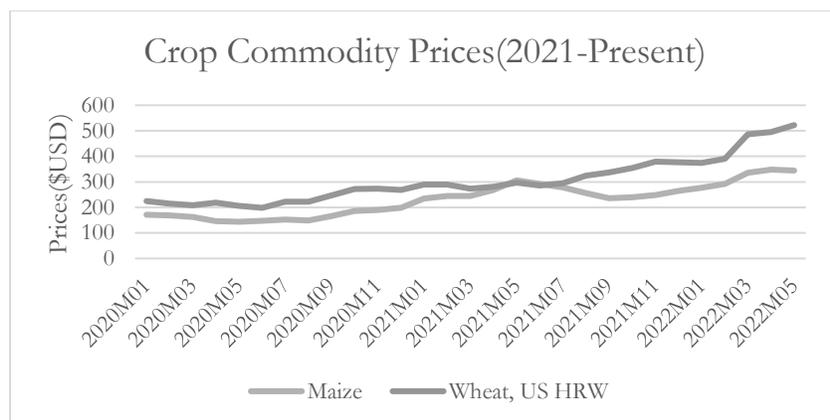

*Fig. 17. Crop Commodity Prices (2021-Present)*

**General Prices**

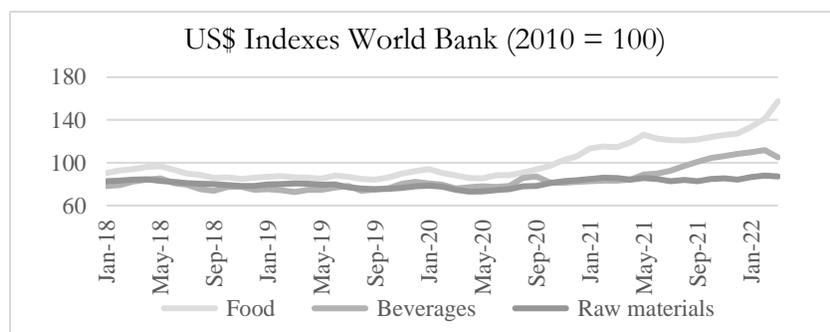

*Fig. 18. US$ Indexes World Bank (2010-Present)*

**Analysis**

From the graphs above we can see the two food commodity prices: Maize and Wheat both follow similar trajectories regarding prices in their respective episodes. In all three episodes both Maize and Wheat Prices increase in prices globally. We can observe that prices have increased generally throughout time. During the 2008 Invasion of Georgia, when maize prices were its highest, it had risen 74% compared to January 2007. Comparatively, when wheat prices were its highest, it had risen 124% compared to January 2007. During the 2014 Annexation of Crimea both wheat and

CRYPTOCURRENCY, SANCTIONS AND AGRICULTURAL PRICES					22maize prices slightly declined but later rose to a comparable level. During the 2022 Invasion of Ukraine both wheat and maize prices increased because of the invasion. Maize prices at its highest increased by 101% compared to January 2021. Similarly when wheat prices were at its highest it had increased by 132% compared to January 2021. The 2022 invasion is still ongoing and prices continuing to change daily. With this in mind, we can make two conclusions. Wheat prices are affected more by Russian invasions and sanctions as compared with maize. Furthermore, the 2022 invasion resulted in the greatest prices. Some possible explanations can be the Covid-19 pandemic and the already rising food prices as seen through the general prices graph.

## Empirical Strategy/Analysis

### Price of Bitcoin and Commodities Correlation

#### Data

Since Bitcoin is the most widely used and popular form of cryptocurrency, this paper obtains live historical close price data from the directly from the Binance Exchange website. Although there are futures markets in all the different countries which we have chosen as samples, the data from these exchanges are either non extractable or unavailable for the commodities which we have chosen. Since many of these exchanges follow similar patterns and fluctuations the paper obtains data from the U.S. futures market gathering data on the commodities Wheat and Corn. Because the goal of this project is to see the impact the Russian Sanctions have on Commodities and how Bitcoin can help, this paper focused on the 2022 Sanctions. The period from January to June 2022 helps suffice for deep analysis when Bitcoin was relatively stable compared to its previous years. Wheat and Corn Futures data was obtained from the U.S. CME Group Derivatives Market.

|  | Wheat Futures | Corn Futures | Bitcoin Price |
|---|---|---|---|
| Wheat Futures | 1 |  |  |
| Corn Futures | 0.923501098 | 1 |  |

|  | Wheat Futures | Corn Futures | Bitcoin Price |
|---|---|---|---|
| Bitcoin Price | -0.381635434 | -0.388289716 | 1 |

*Fig. 19. Wheat, Corn Futures data*

### Analysis

The matrix above is helpful in making clear conclusions regarding the correlation between the three variables. We can conclude that most agricultural commodity prices move in unison with each other. Moreover both Wheat and Corn are affected by the Ukraine Russian Sanctions, so they follow a similar pattern with a correlation of 923501098. On the other hand Bitcoin Prices are not much correlated with either Corn or Wheat Futures prices and have an inverse relationship making them optimal investments. The correlation between Bitcoin and Wheat prices is inverse with a relationship of -0.381635434 and is statistically significant with a p value of less than 0.25. Moreover, the correlation between Bitcoin and Corn prices is inverse with a relationship of -0.388289716 and is statistically significant with a p value of less than 0.25.

**Currency Volatility**

**Data**

Analyzing how volatile a currency is extremely important especially to those in low-income neighborhoods and communities where the money available to spend on external investments is not that high. Therefore, as a more long-term investment, it is important to calculate the volatility to determine the risk of the investments. The data for the exchange rates were obtained from official database whereas the Bitcoin data was obtained from Binance.

|  | Bitcoin | Bangladesh | El Salvador | Iran | Nigeria | Philippines | South Africa |
|---|---|---|---|---|---|---|---|
| Volatility | 0.048665889 | 0.004105435 | 0.001900782 | 0.00155465 | 0.002529668 | 0.003151545 | 0.009713203 |
| Annualized Volatility | 0.772547042 | 0.065558541 | 0.03035306 | 0.024825778 | 0.040395565 | 0.050326143 | 0.155107418 |

*Fig. 20.Volatility*

### Analysis

To obtain the volatility of the currency, exchange rate data shown in $USD was obtained from January 2022 to June 2022. This selection sample was selected to analyze the impact of the 2022 Russian Sanctions. Based on this selection sample, the percent change per date was calculated by performing the following calculation:

$$\hat{U} = (U_i/U_{i-1}) - 1$$

After performing this calculation for the entire historical dataset, the standard deviation of the $\hat{U}$ sample is calculated. This value provides us with the sample volatility. To calculate the annualized volatility the following equation is calculated.

$$Anualized\ Volatiity = \sqrt{225} \times Volatilty$$

225 is the approximate amount of trading days in a year helping provide an annualized value which was factored in when making the initial calculations. Based on this analysis we can conclude that Bitcoin prices are more volatile compared with traditional currencies making them risky investments for low-income communities in the short run.

## Regression of Bitcoin Transactions on Wheat Prices

### Data

Finding the impact Bitcoin Transaction have on Food Prices is important to consider with its implications in a growing market share. This paper therefore implements the Instrumental Variables approach to the problem. Data was collected from the time period January to June 2022 through the database Blockchain.com. Wheat price data was obtained from the U.S. CME Group Derivatives Market.

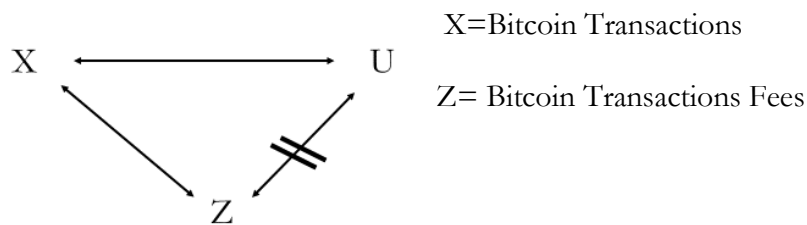

X=Bitcoin Transactions

Z= Bitcoin Transactions Fees

*Fig. 21. Bitcoin Transaction vs Transaction Fees*

Regressing Transaction fees(Z) on Bitcoin transactions(x) provides us with a statistically significant result as the P value is less than 0.25.

|  | Coefficients | Standard Error | T Stat | P-value |
|---|---|---|---|---|
| Intercept | 252481.0835 | 4996.686679 | 50.52970093 | 1.16561E-83 |
| Transaction Fees | 7196.437497 | 2576.367952 | 2.793249113 | 0.006060811 |

*Fig. 22.*

This provides us with the equation: $\hat{X} = 7196Z + 252481 + E$

Using these results and the predicted values we regress $\hat{X}$ onto $Y$ to find the causality correlation between the two variables.

|  | Coefficients | Standard Error | T Stat | P-value |
|---|---|---|---|---|
| Intercept | 795.2940041 | 1031.816865 | 0.770770503 | 0.4423331 |
| Transaction Fees | 0.000886508 | 0.003880228 | 0.22846799 | 0.819664947 |

*Fig. 23.*

This result is not statistically significant and provides us with the equation:

Y=0.000886 $\hat{X}$ + 795+U

**Analysis**

Based on this instrumental variable analysis we can assume that Bitcoin Transactions have no correlation with Food Prices as regression is not statistically significant. This is an important statistic because it allows us to learn that these two variables do not influence each other while other currencies are affected by food prices. Because Bitcoin prices are not affected by rising food prices and vice versa, this might be a potential investment opportunity for citizens of middle- and low-class backgrounds who wish to evade the sanction measures. Another observation is that bitcoin transactions vary and are extremely volatile in a day-to-day manner and trends in adoptability can only be seen over large amounts of time. This therefore does not make it a strong variable to test for adoptability especially in the short run. We can assume the relationship will be similar with corn prices as well as corn and wheat future markets are similarly correlated leading both to result in no relationship with cryptocurrency prices. The fact that there is no relationship is extremely striking as it helps provide justification that transactions numbers are following their regular pattern and are not influenced by or influence wheat prices or agricultural prices in general. We can conclude that the rise in prices did not encourage new users to invest in cryptocurrency and trade; it is the same people on the market trading throughout this period.

**Regression of Agricultural Commodity Prices on Bitcoin Transactions Volume**

### Data

Finding how agricultural commodity prices affect bitcoin transaction volumes is an important estimate to take. The data the transaction volume provides is different compared with the number of transactions because it helps us estimate the value traded as opposed to the amount of people trading. With this knowledge, we can estimate how agricultural commodity prices will affect Bitcoin Transaction Volume. Wheat and Corn Futures data was obtained from the U.S. CME Group Derivatives Market. Bitcoin Transaction Volume shows the value in $USD traded per day on the bitcoin blockchain.

### *Wheat*

|               | Coefficients | Standard Error | T Stat   | P-value   |
|---------------|--------------|----------------|----------|-----------|
| Intercept     | 44409.65     | 19.32519       | 2298.019 | 1.1E-284  |
| Wheat Futures | 0.235564     | 0.018505       | 12.72983 | 3.87E-24  |

*Fig. 24.*

With a P value less than 0.25 and a correlation of 0.755312, we can see that the relationship between Wheat and Bitcoin Volume Transactions is strong. This regression results in the equation:

$$Bitcoin\ Transactions\ Volume\ =\ 0.235\ Wheat\ +\ 44409\ +\ E$$

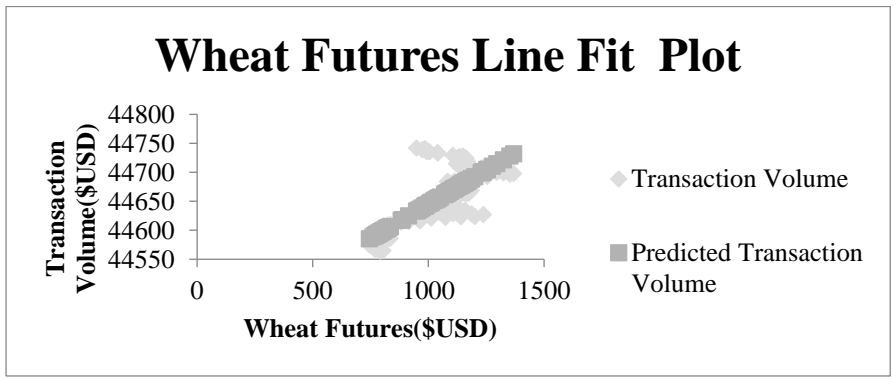

*Fig. 25. Wheat Futures Line Fit Plot*

*Corn*

|  | Coefficients | Standard Error | T Stat | P-value |
|---|---|---|---|---|
| Intercept | 44213.08 | 28.0802 | 1574.528 | 1.2E-264 |
| Corn Futures | 0.60577 | 0.03853 | 15.72219 | 4.07E-31 |

With a P value less than 0.25 and a correlation of 0.818257, we can see that the relationship between Corn Prices and Bitcoin Volume Transactions is strong. This regression results in the equation:

$$Bitcoin\ Transactions\ Volume\ =\ 0.605 Corn\ +\ 44213\ +\ E$$

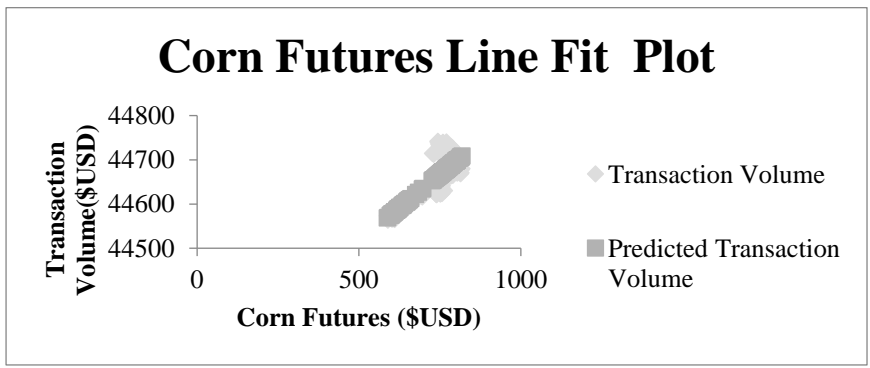

*Fig. 26. Corn Futures Line Fit Plot*

**Analysis**

This analysis helps us justify that as the price of corn and wheat increase the amount of money transacted also increases. From the previous analysis, we concluded that it is the same people on the market trading and the transactions count are not affected by agricultural prices. But the volume increases as the commodity prices increase as evidenced by the regression above. Therefore, we can realize that it is the same few elites who continuously trade on the blockchain but have increased their trading to potentially safeguard their assets during these sanction periods.

**Conclusion**

The results show that both agricultural prices wheat and corn are extremely correlated with each other, suggesting that this relationship might be the same for most agricultural products. Through the graphs we can visualize the direct agricultural commodity price hike because of the Russian Sanctions. Moreover, this study explores the impact these rising prices have on Hunger around the globe. With the increased adoption of cryptocurrencies, this paper shows that Bitcoin Prices are although inversely related not significantly correlated with agricultural commodity prices. Although they are more volatile compared to local currencies, they are not influenced by changes in wheat and corn prices heavily. In each of the 6 countries which we chose, local currencies were stagnant while Bitcoin decreased, and Food Prices increased. This might also make Bitcoin a bad investment for those in low-income communities with less spare income. But since the prices were not correlated, different variables are affecting the Bitcoin price and Bitcoin might be a steady investment for the future. The instrumental variable analysis also shows that Bitcoin Transactions are not correlated with Wheat Prices. Some hypothesized that sanctions would have caused more people to trade using cryptocurrency. But this hypothesis was proved wrong by this analysis. It is therefore safe to conclude that since the two variables did not correlate with each other dramatically, as the number of new users shifting to Bitcoin was not much. But with the analysis on the volume of Bitcoin traded, the results showed that the amount of money sent through Bitcoin increased. This shows that the same elites and middle class who were previously on Bitcoin continued to exchange higher amounts of cash to escape the burdens of sanctions. With these conclusions, low-income communities can leverage this new technology to hedge against sanctions. Similarly to the elites they can facilitate the use of Peer-to-Peer transactions to obtain money from outside countries during times of hard financial hardship.

**Works Cited**

Bonneau, Joseph, et al. "SOK: Research Perspectives and Challenges for Bitcoin and Cryptocurrencies." *2015 IEEE Symposium on Security and Privacy*, May 2015, https://doi.org/10.1109/sp.2015.14.

Boulanger, Pierre, et al. "Russian Roulette at the Trade Table: A Specific Factors CGE Analysis of an Agri-Food Import Ban." *Journal of Agricultural Economics*, vol. 67, no. 2, 2016, pp. 272–291., https://doi.org/10.1111/1477-9552.12156.

Busch, Kristen E, and Paul Tierno. "Russian Sanctions and Cryptocurrency - Congress." *CRS Reports*, Congress Research Services, https://crsreports.congress.gov/product/pdf/IN/IN11920.

Chainalysis. "Geography of Cryptography." *Analysis of Geographic Trends in Cryptocurrency Adoption and Usage*, Chainalysis, 7 Oct. 2021, https://www.chainalysis.com/.

Crozet, Matthieu (01/2020). "Friendly fire: the trade impact of the Russia sanctions and counter-sanctions". Economic policy (0266-4658), 35 (101), p. 97.

Drezner, Daniel W. "The Sanctions Paradox." Economic Statecraft and Economic Relations, 1999, https://doi.org/10.1017/cbo9780511549366.

Drèze, Jean, and Haris Gazdar. "Hunger and Poverty in Iraq, 1991." *World Development*, vol. 20, no. 7, July 1992, pp. 921–945., https://doi.org/10.1016/0305-750x(92)90121-b.

Dudley, Sara, et al. *Evasive Maneuvers: How Malign Actors Leverage Cryptocurrency*. JFQ 92, 2019, https://ndupress.ndu.edu/Portals/68/Documents/jfq/jfq-92/jfq-92_58-64_Dudley-et-al.pdf.

Hinz, Julian, and Evgenii Monastyrenko. "Bearing the Cost of Politics: Consumer Prices and Welfare in Russia." *Journal of International Economics*, vol. 137, July 2022, p. 103581., https://doi.org/10.1016/j.jinteco.2022.103581.

Kuzminov, Ilya, et al. "The Current State of the Russian Agricultural Sector." *EuroChoices*, vol. 17, no. 1, 11 Apr. 2018, pp. 52–57., https://doi.org/10.1111/1746-692x.12184.

Ng, Victor K., and Stephen Craig Pirrong. "Price Dynamics in Refined Petroleum Spot and Futures Markets." *Journal of Empirical Finance*, vol. 2, no. 4, Feb. 1996, pp. 359–388., https://doi.org/10.1016/0927-5398(95)00014-3.

"The Observatory of Economic Complexity Data." *OEC*, Center for Collective Learning, 2022, https://oec.world/.

Rayner, Geof, et al. "Trade Liberalization and the Diet Transition: A Public Health Response." *Health Promotion International*, vol. 21, no. suppl_1, 2006, pp. 67–74., https://doi.org/10.1093/heapro/dal053.

Chainalysis Team, and Author Chainalysis Team. "Crypto Crime Trends for 2022: Illicit Transaction Activity Reaches All-Time High in Value, All-Time Low in Share of All Cryptocurrency Activity." *The 2022 Crpto Crime Report*, Chainalysis, 20 May 2022, https://blog.chainalysis.com/reports/2022-crypto-crime-report-introduction/.

"U.S. Sanctions on Russia." *CRS Reports*, Congressional Research Service, 18 Jan. 2022, https://crsreports.congress.gov/product/pdf/R/R45415.

Working, Holbrook. "The Theory of Price of Storage." The American Economic Review, vol. 39, no. 6, 1949, pp. 1254–62. JSTOR, http://www.jstor.org/stable/1816601. Accessed 10 Jul. 2022.